


\magnification\magstep1
\parskip=\medskipamount
\hsize=6 truein
\vsize=8.2 truein
\hoffset=0.2 truein
\voffset=0.4 truein
\baselineskip=14pt
\tolerance=500


\font\titlefont=cmbx12
 at 10 truept
\font\authorfont=cmcsc10
\font\addressfont=cmsl10 at 10 truept
\font\smallbf=cmbx10 at 10 truept


\newdimen\itemindent \itemindent=13pt
\def\textindent#1{\parindent=\itemindent\let\par=\resetpar%
\indent\llap{#1\enspace}\ignorespaces}

\let\oldpar=\par
\def\resetpar{\oldpar\parindent=0pt\let\par=\oldpar}

\font\ninerm=cmr9 \font\ninesy=cmsy9
\font\eightrm=cmr8 \font\sixrm=cmr6
\font\eighti=cmmi8 \font\sixi=cmmi6
\font\eightsy=cmsy8 \font\sixsy=cmsy6
\font\eightbf=cmbx8 \font\sixbf=cmbx6
\font\eightit=cmti8
\def\eightpoint{\def\rm{\fam0\eightrm}
  \textfont0=\eightrm \scriptfont0=\sixrm \scriptscriptfont0=\fiverm
  \textfont1=\eighti  \scriptfont1=\sixi  \scriptscriptfont1=\fivei
  \textfont2=\eightsy \scriptfont2=\sixsy \scriptscriptfont2=\fivesy
  \textfont3=\tenex   \scriptfont3=\tenex \scriptscriptfont3=\tenex
  \textfont\itfam=\eightit  \def\it{\fam\itfam\eightit}%
  \textfont\bffam=\eightbf  \scriptfont\bffam=\sixbf
  \scriptscriptfont\bffam=\fivebf  \def\bf{\fam\bffam\eightbf}%
  \normalbaselineskip=9pt
  \setbox\strutbox=\hbox{\vrule height7pt depth2pt width0pt}%
  \let\big=\eightbig \normalbaselines\rm}
\catcode`@=11 %
\def\eightbig#1{{\hbox{$\textfont0=\ninerm\textfont2=\ninesy
  \left#1\vbox to6.5pt{}\right.\n@space$}}}
\def\vfootnote#1{\insert\footins\bgroup\eightpoint
  \interlinepenalty=\interfootnotelinepenalty
  \splittopskip=\ht\strutbox %
  \splitmaxdepth=\dp\strutbox %
  \leftskip=0pt \rightskip=0pt \spaceskip=0pt \xspaceskip=0pt
  \textindent{#1}\footstrut\futurelet\next\fo@t}
\catcode`@=12 %


\outer\def\beginsection#1\par{\vskip0pt plus.2\vsize\penalty-150
\vskip0pt plus-.2\vsize\vskip1.2truecm\vskip\parskip
\message{#1}\leftline{\bf#1}\nobreak\smallskip\noindent}


\outer\def\subsection#1\par{\vskip0pt plus.2\vsize\penalty-80
\vskip0pt plus-.2\vsize\vskip0.8truecm\vskip\parskip
\message{#1}\leftline{\it#1}\nobreak\smallskip\noindent}


\newcount\notenumber

\def\note{\advance\notenumber by 1
\footnote{$^{\{\the \notenumber\}}$}}


\tolerance=500


\rightline{Freiburg, THEP-92/33}
\bigskip
{\baselineskip=24 truept
\titlefont
\centerline{ON THE POSSIBILITY OF SPINORIAL QUANTIZATION}
\centerline{IN THE SKYRME MODEL}
}

\vskip 1.1 truecm plus .3 truecm minus .2 truecm

\centerline{\authorfont Domenico Giulini\footnote*{
e-mail: giulini@sun1.ruf.uni-freiburg.de}}
\vskip 2 truemm
{\baselineskip=12truept
\addressfont
\centerline{Fakultt f\"ur Physik,
Universit\"at Freiburg}
\centerline{Hermann-Herder Strasse 3, D-W-7800 Freiburg, Germany}
}
\vskip 1.5 truecm plus .3 truecm minus .2 truecm

\centerline{\smallbf Abstract}
\vskip 1 truemm
{\baselineskip=12truept
\leftskip=3truepc
\rightskip=3truepc
\parindent=0pt

{\eightpoint
We consider the configuration space of the Skyrme model and give a simple
proof that loops generated by $2\pi$-rotations are contractible in the even-,
and non-contractible in the odd-winding-number sectors.
\par}}

\beginsection Introduction

As is well known, the Skyrme model and related models allow quantizations
with odd-half angular momentum due to the non-contractibility of some loops
in the configuration space (represented by an infinite dimensional
mapping-space).
Standard expositions, treating the quantization of the collective
degrees of freedom in the unit winding-number sector, usually argue only
within the (finite dimensional) subspace of collective coordinates
(e.g. [Ba][Na-Wi]). But non-contractibility of certain loops in a subspace
does not imply non-contractibility when allowed to be deformed outside this
space. Valid proofs for the unit winding-number sector have been given first
in [Wi] and later in [Wi-Zv][Sk][Pa-Tz]. From [Pa-Tz] one might get the
impression that this is all there is to prove. But this is not the case and
other winding number sectors require in principle a separate treatment.
To fill this gap, this article presents a proof for the statement made in
the abstract. We tried to design it as elementary and self-contained as
possible.

\beginsection Section 1: General Setting

Let us start by a general consideration of a configuration space $Q$ together
with a right $SO(3)$- action which we write
$$\eqalign{
R: (SO(3)\times Q) & \rightarrow Q               \cr
            (g,q) & \mapsto R_g(q)=q\cdot g\,.  \cr}
\eqno{(1)}
$$
A curve $\gamma :[0,1]\rightarrow SO(3)$ then defines the flow
$\phi_q(t)=q\cdot\gamma(t)$ in $Q$. We are interested in the case
where $\gamma$ is a closed loop in $SO(3)$ representing a
non-contractible loop. Since $\pi_1(SO(3))\cong Z_2$, we want
$[\gamma]$ (denoting its homotopy class) to be the generator of $Z_2$.
In particular, any $\gamma$ describing a $2\pi$-rotation about a fixed axis
will do. Now, the question is whether the loop at $q$,
$\phi_q: t\mapsto\phi_q(t)$, is contractible in $Q$. Clearly, a necessary
condition is that $\pi_1(Q,q)$ posesses a $Z_2$ subgroup ($4\pi$-rotations
are necessarily contractible). If $q,p\in Q$ are connected by a path
$c :\,t\mapsto c(t)$, $c(0)=q$, $c(1)=p$, there is the
standard isomorphism $c^{\#}:\,\pi_1(Q,p)\rightarrow\pi_1(Q,q)$ (see
[St] \S 16 ).
The one-parameter family of loops,
$$
\lambda_s:\, t\mapsto\cases{
c(3st)               & for $t\in [0,{1\over 3}]$          \cr
\phi_{c(s)}(3t-1)    & for $t\in [{1\over 3},{2\over 3}]$ \cr
c(3s(1-t))           & for $t\in [{2\over 3},1]\,,$       \cr}
\eqno{(2)}
$$
define a continuous deformation between representatives of $[\phi_{q}]$
and $c^{\#}[\phi_{p}]$, showing that $[\phi_q]$ is trivial in $\pi_1(Q,q)$,
if and only if $[\phi_p]$ is trivial in $\pi_1(Q,p)$. Since we certainly
assume the space $Q$ to be locally pathwise connected, the path-components
coincide with the connected components, and one has the

\proclaim
Lemma 1. Contractibility of $\phi_q$ depends only on the connected
         component of $q$ in $Q$ and not on the choice of $q$ whithin
         that component.

If we decompose $Q$ into its connected components $Q_n$, we can now ask the
question, whether for a given $SO(3)$-action $R$ on $Q$ the homotopy class
of loops $[\phi_n]$ (due to Lemma 1 we omit indicating the basepoint) is
contractible or not. Following [Fi-Ru], we make the following

\proclaim
Definition 1. The configuration space $Q$ is said to allow for spinorial
              states in the sector $Q_n$ with respect to the action $R_g$
              of the spatial group $SO(3)$, if and only if the loop
              $t\mapsto R_{g_t}(q)$ is non-contractible, for $t\mapsto g_t$
              any representative of the generator of $Z_2\cong\pi_1(SO(3))$,
              and any $q\in Q_n$.

If the components $Q_n$ are mutually homeomorphic, as it is for example
the case if $Q$ is a topological group, the homeomorphisms
$h_{mn}:\, Q_n\rightarrow Q_m$ then also define isomorphisms on all
homotopy groups, hence, in particular, on the fundamental groups
$\pi_1(Q_n)$. This implies

\proclaim
Lemma 2. The loop $\phi_n$ is contractible in $Q_n$, if and only if the
         loop $h_{mn}\circ\phi_n$ is contractible in $Q_m$.

It is important to note that contractibility of $h_{mn}\circ\phi_n$
has generally nothing to do with contractibility of $\phi_m$, the first loop
being just a homeomorphic image in $Q_m$ of the loop $\phi_n$ in $Q_n$,
which might not reflect how the group acts in $Q_m$.
The use of Lemma 2 lies in the fact that for standard applications
there is a distinguished component, say $Q_0$, which is analytically more
accessible, so that all the constructions can be performed on $Q_0$ using
$h_{0n}$.

\beginsection  Section 2: The $SU(2)$ Skyrme Model Admits Spinorial States

In this section we continue our notation from the previous section.
In particular, $q,p,..$ still denote points in the configuration space
which is now a mapping space so that e.g. $q$ denotes both, a point in $Q$
and a map. Integer subscripts usually refer to the connected component
the point lies in, i.e., $q_n\in Q_n$.

The $SU(2)$ Skyrme model is described by a map $q$ from $R^3$
into $SU(2)\cong S^3$, dynamically stabilized by adding higher order
terms in the first derivatives of the field variables. A geometric
interpretation is given in [Ma]. Finite energy requires the map to be
extendable to the compactification $S^3$ of $R^3$ by adding a point,
called $\infty$, at infinity, which must be mapped to the identity element
$e$ in $SU(2)$. The configuration space is then the space of basepoint
preserving maps
$$
Q = \{q : \, (R^3 \cup\infty,\infty)\rightarrow (SU(2),e)\}
\eqno{(3)}
$$
whose path-components are just the homotopy classes
$$
\pi_0(Q) = \pi_3 (S^3) = Z\,.
\eqno{(4)}
$$
Although the domain- and target-space can both be identified with $S^3$,
we notationally separate them by using $R^3\cup\infty$ or $S^3$ for the
domain-, and $SU(2)$ for the target-space, also stressing the additional
group structure of the latter.

A static configuration minimizes the energy functional which is bounded
below by a real number proportional to the winding number $W[q]$ of the
map
$$
R^3\cup\infty\mathop{\longrightarrow}^{q} SU(2)\,,
\eqno{(5)}
$$
which can be identified with the integer appearing in (4). The winding
number has the analytic expression [Bo-Se]
$$
W[q]={1\over 24{\pi}^2}\int_{S^3}\,{\rm tr}\,(q^{-1}\,dq)^3\,,
\eqno{(6)}
$$
and can be considered as the piecewise constant function
$W: Q\rightarrow Z$, $W[q]=n\Leftrightarrow q\in Q_n$.
Since the target space is a topological group, the space $Q$ can
also be given the structure of a topological group by pointwise
multiplication, i.e. $(q_1q_2)(x):=q_1(x)q_2(x)\,\forall x\in S^3$, and
a suitable choice of topology on $Q$ (e.g. compact-open or finer).
{}From expression (6) it easily follows that
$$
W[q_1q_2]=W[q_1]+W[q_2]\,,
\eqno{(7)}
$$
so that we can define homeomorphism from $Q_n$ to $Q_m$ by right
(conventionally) multiplication with an element $p_{m-n}\in Q_{m-n}$
$$\eqalign{
h_{mn}:\quad Q_n & \rightarrow Q_m   \cr
             q_n & \mapsto h_{mn}(q_n):
             =q_np_{m-n}\qquad\forall q_n\in Q_n\,.\cr}
\eqno{(8)}
$$
In order to find the fundamental group of $Q$ (i.e. of each component $Q_n$),
we only need to find $\pi_1(Q_0,b)$ for some basepoint $b$. We may therefore
choose $b$ to be the constant map $S^3\mapsto e\in SU(2)$.
On the other hand, a loop in $Q_0$ at $b$ defines a one parameter family
(labelled by the unit intervall $I$) of maps from the solid
three-dimensional cube $K$ to $SU(2)$, starting and ending at $b$, and
such that the boundary $\partial K$ of $K$ is mapped to $e\in SU(2)$:
$$\eqalign{
                     & \alpha:\,I\times K\rightarrow SU(2)       \cr
   \hbox{where}\quad & \alpha (0\times K)=\alpha (1\times K)=e   \cr
   \hbox{and}  \quad & \alpha (I\times \partial K) = e  \,.      \cr}
   \eqno{(9)}
$$
But this is a map from the solid four-dimensional cube
$I\times K$ to $SU(2)$ whose boundary is mapped to $e\in SU(2)$.
Homotopic loops in $Q$ at $b$ thus define homotopic maps from the four-cube
with the boundary mapped into $e$, and vice versa. Therefore, there is an
isomorphism:
$$
\pi_1(Q,b)\cong \pi_4(SU(2),e)\cong Z_2\,.
\eqno{(10)}
$$
Since this group is abelian, we need not indicate the
basepoint\note{Note that there are generally no canonical
isomorphisms of the fundamental groups at different basepoints. Isomorphisms
are only defined up to inner automorphisms. A canonical identification
therefore exists for abelian groups.} and we can unambiguously write
$$
\pi_1(Q)=\pi_1(Q_n)=Z_2\,.
\eqno{(11)}
$$

Next we need to specify $R$, the $SO(3)-$action on $Q$. It has the obvious
action on the spatial $R^3$ and extends to the compactification
$S^3=R^3\cup\infty$ by fixing $\infty$. Denoting the $SO(3)$-matrix by
$M$ and by $Mx$ the action of $M$ on $x\in R^3$, we define $R$ via
$$
[R_Mq](x):=q(Mx)\quad\forall x\in R^3\,.
\eqno{(12)}
$$
A closed curve $\gamma$ in $SO(3)$, represented by the one-parameter family of
matrices $M_t$, then defines the free (i.e. unbased) homotopy class $[\phi_n]$
of loops in $Q_n$ represented by the loops $\phi_{q_n}$ in $Q_n$ given by
$$
\phi_{q_n}(t):= q_n\circ M_t\,.
\eqno{(13)}
$$
Right multiplication by
$[q_n]^{-1}$\note{If $i$ denotes the map $g\mapsto i(g):=g^{-1}$ on $SU(2)$,
then we write $i\circ q=:[q]^{-1}$ or $(i\circ q)(x)=:[q(x)]^{-1}$.}
then defines a loop $\varphi_n$ in $Q_0$ based at the constant map
$R^3\cup\infty \mapsto e$:
$$
\varphi_n(t)=\phi_{q_n}(t)[q_n]^{-1}\quad\hbox{or}\quad
\varphi_n^t(x)=q_n(M_t x)[q_n(x)]^{-1}\,.
\eqno{(14)}
$$
According to what has been said above, the $W=n$ sector admits spinorial
states, if and only if $\varphi_n$
is non-contractible, or, as expressed in (10), if and only if
$\varphi_n$ generates $Z_2=\pi_4(SU(2))$.

\subsection The Truncated Model

Before proceeding with the general case,
let us very briefly review the standard arguments for the truncated model
in the $W=1$ sector. Here one introduces collective coordinates, $A(t)$,
around the energy minimizing map $m\in Q_1$ in the $W=1$ sector, using
the ansatz
$$\eqalignno{
                 q(t,x) & = A (t) m(x) A^{-1}(t)\,,  & (15) \cr
\hbox{where}\quad  m(x) & = \cos [f(r)] + i\vec n\cdot\vec\tau \
                                     \sin [f(r)] \,, & (16) \cr}
$$
where, $\vec\tau$ denotes the triplet of Pauli matrices and
$\vec n=\vec x/\vert\vec x\vert$. In this way one truncates the
configuration space $Q$ to
$$
Q_{\rm tr} = SU(2) / Z_2 \cong RP^3\,,
\eqno{(17)}
$$
since $A(t)$ and $-A(t)$ are to be identified. It is obvious that a spatial
rotations with $SO(3)$ matrix $M$ and corresponding $SU(2)$ covering
element $S$ obeys
$$
q(t,M x) = A S m S^{-1} A^{-1} \,,
\eqno{(18)}
$$
i.e., rotations act on $Q_{tr}$ by right multiplication.
A 1-parameter family $S_t$ generating a $2\pi$-rotation therefore generates
$Z_2=\pi_1 (Q_{\rm tr})$. Hence the truncated model admits spin in the
sense of the definition given above. But clearly, this does not imply the
same statement in the full theory.

\subsection The General Case

Let us first consider the sector $Q_1$. As basepoint we choose the map
$q_1\in Q_1$ corresponding to the inverse stereographic projection:
$$\eqalign{
q_1(\vec x) = & {1 - r^2 \over 1+r^2}+i\vec\tau\cdot\vec n(\vec x)
            {2r \over r^2 + 1}
            \qquad(\hbox{where}\quad\vec n(\vec x)=\vec x/r)    \cr
         = & : \,a + i\vec\tau\cdot \vec a \,,                  \cr}
\eqno{(19)}
$$
where the last line just introduces $a$ and $\vec a$ as abreviations of
the expressions above. As loop $\gamma$ in $SO(3)$ we choose
$$
M:\,[0,2\pi]\ni t\mapsto
M_t = \pmatrix{\cos t&\sin t&0\cr -\sin t&\cos t&0\cr0&0&1&\cr}\,.
\eqno{(20)}
$$
Its $SU(2)$-lift is given by
$$
S_t=\exp\left({i\over 2}\tau_3 t \right)\,,
\eqno{(21)}
$$
so that according to (14) and (19)
$$
\varphi^t_{1}(x)  = q_1(M_tx)[q_1(x)]^{-1}
                  = S_tq_1(x)S_t^{-1}[q_1(x)]^{-1}  \,.
\eqno{(22)}
$$
Inserting (19), we obtain
$$\eqalignno{
\varphi_1^t & = (a+i\vec\tau\cdot M_t\vec a)(a-i\vec\tau\cdot\vec a)
              = : \, (a'+i\vec\tau\vec a')                                \cr
            & = a^2+\vec a\cdot M_t\vec a+i\tau_3 (a^2_1 + a^2_2) \sin t +
          {\rm terms} \propto \ \tau_1\, {\rm and}\,\tau_2\,.
      & (23)                                                          \cr}
$$
For $0 \leq t \leq \pi$ this maps onto the hemisphere $a'_3 \geq 0$,
for $\pi\leq t\leq2\pi $ onto the hemisphere $a'_3 \leq 0$, and for
$t = \pi$ onto the equator $a'_3= 0$. It is therefore a suspension
(defined by these conditions) of the ``equator-map''
$$
\varphi_1^{t = \pi} : \ S^3\rightarrow S^2:= \{ (a'_1,a'_ 2,a')\ \vert\
                  a'^2_1 + a'^2_2 + a'^2 = 1\}\,,
\eqno{(24)}
$$
which, using expressions (21-22), is given by
$$
\varphi_1^{\pi} = S_{\pi} q_1 S^{-1}_{\pi} [q_1]^{-1}
          = \tau_3 q_1 \tau_3 [q_1]^{-1}\,.
\eqno {(25)}
$$

Now, there is a special map from $S^3$ to $S^2$, called the Hopf map, which
generates $\pi_3(S^2)=Z$. It is given by the projection map
$S^3\rightarrow S^2$ of the Hopf bundle that fibres $S^3$ by $S^1$
over $S^2$. Identifying $SU(2)\cong S^3$, it is simply given
by\note{In Euler angles on $S^3$ and polar angles on $S^2$ this corresponds
to $(\psi, \theta, \varphi)\rightarrow (\theta, \varphi)$.}
$$
h(g)=g\tau_3 g^{-1}=:\vec h(g)\cdot \tau\,, \quad\hbox{where}\quad
                                            \vec h\cdot\vec h=1\,,
\eqno{(26)}
$$
so that (25) now reads
$$
\varphi^\pi_1 = \tau_3 h\circ q_1 = \tau_3\vec\tau\cdot\vec h\circ q_1 \,,
\eqno{(27)}
$$
which shows that this is the Hopf map onto the two-Sphere in $(a'_1,a'_2,a')$
coordinates, composed with a rotation in the $a'_1$-$a'_2$ plane about an
angle ${\pi \over 2}$.  In particular, it is homotopic to the Hopf map.
We now use the standard result in homotopy theory (see [St] 21.4, 21.6) that
any suspension of the Hopf map, and therefore all its homotopies, generate
$Z_2=\pi_4 (S^3)$. This then proves the existence of spinorial quantizations
in the sector $Q_1$.

For $Q_n$ we can proceed as above by choosing as
basepoint $q_n\in Q_n$ a slight modification of the ansatz (19), in which
$\vec n(\vec x)$ is replaced by a map $\vec s_n:=w_n\circ\vec n$, where
$w_n$ is a winding number $n$ map of the two-sphere onto itself. In
spherical polar coordinates $(\theta, \varphi)$ ($\theta$ being as usual
the angle to the $z$-axis) it is defined by
$$
\eqalign{
w_n:\,  S^2      & \rightarrow S^2            \cr
(\theta,\varphi) & \mapsto w_n(\theta, \varphi):
                   =(\theta, n\varphi)\,,     \cr}
\eqno{(28)}
$$
so that the basepoint $q_n$ is now given by
$$\eqalign{
q_n(\vec x)= & \,{1 - r^2 \over 1+r^2}+i\vec\tau\cdot\vec s_n(\vec x)
               {2r \over r^2 + 1}                  \cr
           = &:\,a_n + i\vec\tau\cdot \vec a_n\,.  \cr}
\eqno{(29)}
$$
It is obvious that $q_n\in Q_n$ (i.e. it satisfies $W[q_n]=n$).
Choosing the same rotation map $M_t$ as in the $n=1$ case, we now have the
crucial property
$$
\vec s_n(M_t\vec x)=M_{nt}\vec s_n(\vec x)\,,
\eqno{(30)}
$$
which follows readily from (20) (in polar angles: $M_t(\theta,\varphi)=
(\theta, \varphi+ t)$) and (28). Mapping back the loop $q^t_n$ to a loop
$\varphi_n^t$ in $Q_0$ via right $[q_n]^{-1}$-multiplication, we obtain
$$\eqalignno{
\varphi_n^t & = (a_n+i\vec\tau\cdot M_{nt}\vec a_n)(a_n-i\vec\tau\cdot\vec a_n)
              = : \, (a_n'+i\vec\tau\vec a_n')                         \cr
            & = a_n^2+\vec a_n\cdot M_{nt}\vec a_n+i\tau_3 (a^2_{n1} +
            a^2_{n2})\sin nt + {\rm terms} \propto \ \tau_1\, {\rm and}\,
            \tau_2\,.
            & (31)                                                     \cr}
$$
As above, we conclude that the partial loop for $0\leq t\leq 2\pi/n$
represents the generator of $Z_2=\pi_4(SU(2))$. The $n$-fold loop
therefore represents the generator for $n=$ odd, and the trivial element
for $n$=even. We have thus proven the

\proclaim
Theorem. The Skyrme model admits spinorial states in the sectors of odd winding
number, and no spinorial states in the sectors of even winding number.

Due to the topological spin-statistics theorem, proven in [Fi-Ru] (see also
[So] for an elegant formulation), we have at the same time the corresponding
statement for exchangement in all winding number sectors: The Skyrme model
allows for fermionic quantizations in the odd winding number sectors, and
bosonic quantizations only in the even winding number sectors.

\vfill\eject

\beginsection{References}

{\parskip=0.1truecm

\item{[Bo-Se]}  Bott, R., Seeley, R.: Comm. Math. Phys., {\bf 62}, 235
                (1978)

\item{[Fi-Ru]}  Finkelstein, D., Rubinstein, J.: Jour. Math. Phys. {\bf 9},
                1762 (1968)

\item{[Sk]}     Skyrme, T.H.R.: Jour. Math. Phys., {\bf 12}, 1735-1743 (1971)

\item{[St]}     Steenrod, N.: The Topology of Fibre Bundles. Princeton
                University Press 1974 (ninth printing).

\item{[So]}     Sorkin, R.D.:Comm. Math. Phys. {\bf 115}, 421-434 (1988)

\item{[Wi]}     Williams, J.G.: Jour. Math. Phys., {\bf 11}, 2611-2616 (1970)

\item{[Wi-Zv]}  Williams, J.G., Zvengrovski, P.: Int. Jour. Theo. Phys.,
                {\bf 16}, 755-761 (1977)

}
\end